\documentclass[prl, aps, superscriptaddress, amsmath, amssymb, reprint, floatfix]{revtex4-1}

\usepackage{graphicx}
\usepackage{textcomp}
\usepackage{color}

\usepackage{todonotes}

\begin{document}

\title{Dynamic acousto-mechanical control of a strongly coupled photonic molecule}

\author{Stephan Kapfinger}
\affiliation{Lehrstuhl f\"{u}r Experimentalphysik 1 and Augsburg Centre for Innovative Technologies (ACIT), Universit\"{a}t Augsburg, Universit\"{a}tstr. 1, 86159 Augsburg, Germany}
\affiliation{Nanosystems Initiative Munich (NIM), Schellingstr. 4, 80799 M\"{u}nchen, Germany}

\author{Thorsten Reichert}
\affiliation{Walter Schottky Institut and Physik Department, Technische Universit\"{a}t M\"{u}nchen, Am Coulombwall 4, 85748 Garching, Germany}
\affiliation{Nanosystems Initiative Munich (NIM), Schellingstr. 4, 80799 M\"{u}nchen, Germany}

\author{Stefan Lichtmannecker}
\affiliation{Walter Schottky Institut and Physik Department, Technische Universit\"{a}t M\"{u}nchen, Am Coulombwall 4, 85748 Garching, Germany}
\affiliation{Nanosystems Initiative Munich (NIM), Schellingstr. 4, 80799 M\"{u}nchen, Germany}

\author{Kai M\"{u}ller}
\affiliation{Walter Schottky Institut and Physik Department, Technische Universit\"{a}t M\"{u}nchen, Am Coulombwall 4, 85748 Garching, Germany}
\affiliation{Nanosystems Initiative Munich (NIM), Schellingstr. 4, 80799 M\"{u}nchen, Germany}
\affiliation{E. L. Ginzton Laboratory, Stanford University, Stanford, California 94305, United States}

\author{Jonathan J. Finley}
\affiliation{Walter Schottky Institut and Physik Department, Technische Universit\"{a}t M\"{u}nchen, Am Coulombwall 4, 85748 Garching, Germany}
\affiliation{Nanosystems Initiative Munich (NIM), Schellingstr. 4, 80799 M\"{u}nchen, Germany}

\author{Achim Wixforth}
\affiliation{Lehrstuhl f\"{u}r Experimentalphysik 1 and Augsburg Centre for Innovative Technologies (ACIT), Universit\"{a}t Augsburg, Universit\"{a}tstr. 1, 86159 Augsburg, Germany}
\affiliation{Nanosystems Initiative Munich (NIM), Schellingstr. 4, 80799 M\"{u}nchen, Germany}
\affiliation{Center for NanoScience (CeNS), Ludwig-Maximilians-Universit\"{a}t M\"{u}nchen, Geschwister-Scholl-Platz 1, 80539 M\"{u}nchen, Germany}

\author{Michael Kaniber}
\affiliation{Walter Schottky Institut and Physik Department, Technische Universit\"{a}t M\"{u}nchen, Am Coulombwall 4, 85748 Garching, Germany}

\author{Hubert J. Krenner}
\affiliation{Lehrstuhl f\"{u}r Experimentalphysik 1 and Augsburg Centre for Innovative Technologies (ACIT), Universit\"{a}t Augsburg, Universit\"{a}tstr. 1, 86159 Augsburg, Germany}
\affiliation{Nanosystems Initiative Munich (NIM), Schellingstr. 4, 80799 M\"{u}nchen, Germany}
\affiliation{Center for NanoScience (CeNS), Ludwig-Maximilians-Universit\"{a}t M\"{u}nchen, Geschwister-Scholl-Platz 1, 80539 M\"{u}nchen, Germany}

\maketitle

{\bf Two-dimensional photonic crystal membranes provide a versatile planar architecture for integrated photonics to control the propagation of light on a chip employing high quality optical cavities, waveguides, beamsplitters or dispersive elements \cite{Notomi_2010}.
When combined with highly non-linear quantum emitters, quantum photonic networks \cite{OBrien_2009,Faraon_2011} operating at the single photon level \cite{Volz_2012} come within reach.
Towards large-scale quantum photonic networks \cite{Hartmann_2006,Yang_2009}, selective dynamic control of individual components and deterministic interactions between different constituents are of paramount importance \cite{Grillet_2010}.
This indeed calls for switching speeds ultimately on the system's native timescales \cite{Shevchenko_2010}.
For example, manipulation via electric fields or all-optical means have been employed for switching in nanophotonic circuits \cite{Vlasov_2005,Tanabe_2007} and cavity quantum electrodynamics studies \cite{Jin_2014,Laucht_2009}.
Here, we demonstrate dynamic control of the coherent interaction between two coupled photonic crystal nanocavities forming a photonic molecule \cite{Bayer_1998,Chalcraft_2011}.
By using an electrically generated radio frequency surface acoustic wave we achieve optomechanical tuning \cite{Fuhrmann_2011}, demonstrate operating speeds more than three orders of magnitude faster than resonant mechanical approaches \cite{Li_2014}. Moreover, the tuning range is large enough to compensate for the inherent fabrication-related cavity mode detuning.
Our findings open a route towards nanomechanically gated protocols \cite{Blattmann_2014,Schuelein_2015}, which hitherto have inhibited the realization in all-optical schemes \cite{Sato_2012,Bose_2014}.
}

\begin{figure*}[htb]
\centering
\includegraphics[width=1.7\columnwidth]{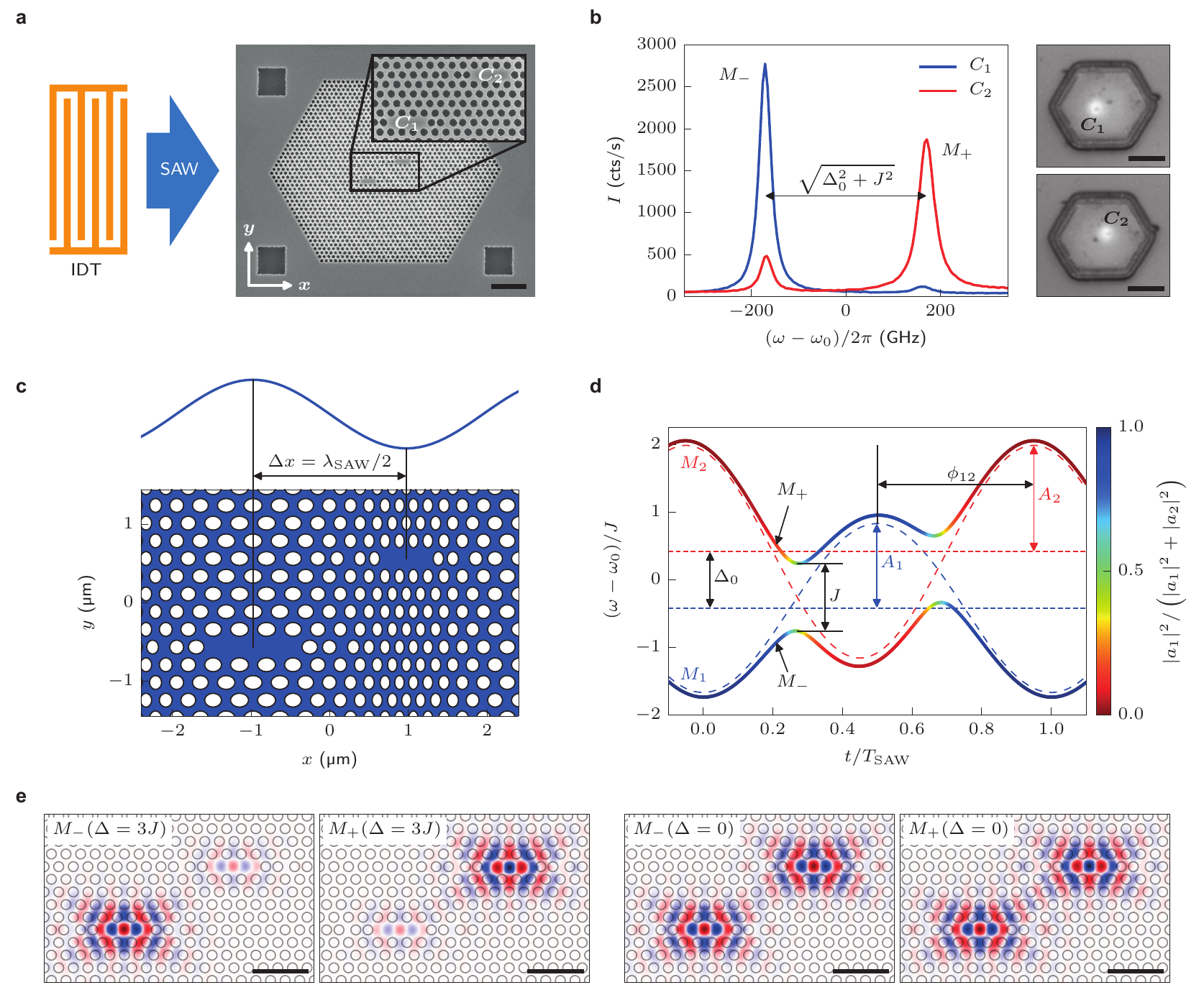}
\caption{
Photonic molecule with acousto-mechanical tuning mechanism.
(a)~SEM image of the photonic molecule and schematic of the device layout.
The IDT launches a SAW which propagates across the photonic molecule.
Scale bar, 2~\textmu m (IDT not to scale).
(b)~PL spectra showing the two fundamental modes of the photonic molecule and optical microscope images of the excitation laser spot.
The blue and the red curve were obtained by exciting on cavity $C_1$ and $C_2$, respectively.
Due to a finite static detuning $\Delta_0>J$, the modes are confined within their respective cavities.
The center frequency $\omega_0$ is $2\pi\times\mathrm{305~THz}$.~  The cavities can be spatially resolved in our optical setup, allowing selective excitation of the individual cavities.
Scale bar, 5~\textmu m.
(c)~Acousto-mechanical tuning.
The resonance frequencies of the cavities are modulated by the SAW induced strain field (amplitude exaggerated).
Setting half the SAW wavelength equal to the cavity separation, the cavities can be tuned relative to each other, giving rise to a time dependent detuning $\Delta(t)=\Delta_0+\Delta_\mathrm{mod}\sin(\omega_{\rm SAW} t)$.
(d)~Evolution of the modes over one SAW period.
The dashed lines represent the single cavity modes $M_{1}$ and $M_{2}$, and the solid lines the normal modes $M_{+}$ and $M_{-}$.
The color scale corresponds to the decomposition into single cavity modes, with green denoting the symmetric and antisymmetric superposition on resonance.
The static detuning $\Delta_0$ and the coupling strength $J$ are intrinsic parameters of the photonic molecule.
The tuning amplitudes $A_{1,2}$ and the phase shift $\phi_{12}$ are determined by the amplitude and wavelength of the SAW, respectively.
(e)~FDTD simulations of the mode profiles ($E_y$).
For the case of detuned cavities ($\Delta=3J$), the modes are localized within the individual cavities.
On resonance ($\Delta=0$), one obtains the symmetric (bonding) and the antisymmetric (antibonding) mode.
}
\label{fig1}
\end{figure*}

The on-chip photonic molecule (PM) studied here consists of two L3-type missing hole cavities, labeled $C_1$ and $C_2$, defined in a two-dimensional photonic crystal membrane.
Its static coupling strength $J$ is known to depend exponentially on the separation between the two cavities \cite{Chalcraft_2011}.
In our sample the cavities are offset symmetrically by $d=5$ lattice constants $(a=260\,{\rm nm})$ along each primitive direction of the photonic crystal (PC) lattice (see Supplementary information), as can be seen in the scanning electron micrograph in Fig. \ref{fig1} (a).
The finite coupling strength $J$ leads to the formation of two normal modes; a bonding mode $M_-$ and an antibonding mode $M_+$.
Note that the mode indices in $M_\pm$ refer to the respective normal mode frequencies $\omega_\pm$ and, thus, are opposite to the \emph{spatial} symmetry of the mode. In the most general case, the two cavities $C_1$ and $C_2$ exhibit resonance frequencies $\omega_1$ and $\omega_2$, split by a finite detuning $\Delta=\omega_2-\omega_1$.
Thus, the normal mode frequencies can be expressed by
\begin{equation}
\omega_\pm=\omega_0\pm\frac{1}{2}\sqrt{\Delta^2+J^2}
\label{coupledmodes}
\end{equation}
with $\omega_0=(\omega_1+\omega_2)/2$ being the centre frequency.
In Fig. \ref{fig1} (b) we present two emission spectra of embedded quantum dots (see methods) recorded for spatially exciting either $C_1$ (blue) or $C_2$ (red) of a typical PM, denoted as PM1.
For PM1, each cavity exhibits a distinct \emph{single} mode, split by $\omega_\pm=\omega_0\pm\frac{1}{2}\sqrt{\Delta_0^2+J^2}$, with $\Delta_0$ being the \emph{static} detuning. 
Therefore, these as-fabricated and nominally identical cavities are efficiently decoupled due to inevitable imperfections during nanofabrication.
To overcome this fundamental limitation and achieve dynamic control of the individual cavity resonances we employ an optomechanical approach using radio frequency (rf) surface acoustic waves (SAWs) \cite{Lima_2005,Fuhrmann_2011}.
As indicated in the schematic illustration of Fig. \ref{fig1} (a), these quasi-monochromatic acoustic phonons are generated by an interdigital transducer (IDT) (see methods), and propagate along the $x$-axis of the PM and dynamically deform the individual cavities.
The amplitude of this mechanism, $A/2\pi > 250\,{\rm GHz}$, exceeds both the static detuning, $\Delta_0$, and the coupling strength, $J$, which enables us to dynamically tune the PM completely in and out of resonance.  
For our chosen geometric arrangement, the cavities are separated by a distance $\Delta x$ along the SAW propagation direction, the modulations of $\omega_1$ and $\omega_2$ are phase shifted by $\phi_{12}=\Delta x\cdot\omega_{\rm SAW}/c_{\rm SAW}$, with $\omega_{\rm SAW}$ and $c_{\rm SAW}$ being the SAW angular frequency and speed of sound, respectively.
We deliberately set the acoustic wavelength, $\lambda_{\rm SAW}$, such that it is commensurate with the spatial separation of the two cavity centers $2\Delta x=\lambda_{\rm SAW}$.
As illustrated in the schematic in Fig. \ref{fig1} (c), for the selected phase, the maximum (minimum) of the SAW expands $C_1$  (compresses $C_2$), which in turn gives rise to a red (blue) shift of their respective resonances \cite{Fuhrmann_2011}.
The propagation of the sound wave along the $x$-axis of the PM leads to a \emph{time-dependent detuning} of the two cavities
\begin{equation}
\Delta(t)=\Delta_0+\Delta_\mathrm{mod}\sin(\omega_{\rm SAW} t).
\label{modulation}
\end{equation}
The amplitude of this modulation, $\Delta_\mathrm{mod}=\sqrt{A_1^2+A_2^2-2A_1A_2\cos(\phi_{12})}$, depends on the amplitudes of the modulations of the individual cavities, $A_1$ and $A_2$, and is maximum for $\phi_{12}=180\textrm{\textdegree}$.
For $\Delta_{\rm mod}>\Delta_0$, the detuning passes through zero, which results according to equation (\ref{coupledmodes}) in an avoided crossing of the normal modes.
In Fig. \ref{fig1} (d), the resulting temporal evolutions of the normalized mode frequencies $(\omega-\omega_0/J)$ obtained using equations (\ref{coupledmodes}) and (\ref{modulation}) are evaluated over one acoustic cycle of $T_{\rm SAW}$ for a typical set of experimentally achievable parameters ($J=1.2\cdot \Delta_0$, $A_1=1.5\cdot \Delta_0$, $A_2=1.9\cdot \Delta_0$ and $\phi_{12}=162\textrm{\textdegree}$).
The dashed lines show the time-evolution of the non-interacting $(J=0)$ modes $M_1$ (blue) and $M_2$ (red).
In strong contrast, the normal modes $M_+$ and $M_-$ start with initially $M_1$-like (blue) and $M_2$-like (red) single-cavity character and develop to fully mixed symmetric (bonding) and antisymmetric (antibonding) character (green) at resonance. At this point, the coupling strength $J$ of the PM can be deduced directly from the splitting of the avoided crossing. After traversing through the avoided crossing, the initial character of the modes is exchanged, with $M_+$ and $M_-$ possessing $M_2$-like (blue) and $M_1$-like character, respectively.
Over the duration of one full acoustic cycle the two modes are brought into resonance at two distinct times, giving rise to two  avoided crossings, restoring the initial configuration.
We performed full finite difference time-domain (FDTD) simulations to confirm and in particular quantify the dynamic coupling.
We evaluate the calculated profiles of the electric field component in the plane of the photonic crystal membrane and perpendicular to the SAW propagation $(E_y)$ in Fig. \ref{fig1} (e).
In the two left panels, the detuning is set to three times the coupling strength $(\Delta=3J)$ and, thus, the two modes remain well localized in one of the two cavities.
For the resonance case, $\Delta =0$, shown in the two right panels of Fig. \ref{fig1} (e), the $E_y$ exhibit the characteristic symmetric and antisymmetric superpositions for the $M_-$ and $M_+$ modes, respectively.
For our sample geometry the FDTD simulations predict $J_\mathrm{sim}/2\pi=170\,{\rm GHz}$.
This value is set by the chosen separation, i.e. barrier thickness, between the two cavities.

\begin{figure*}[htb]
\centering
\includegraphics[width=1.7\columnwidth]{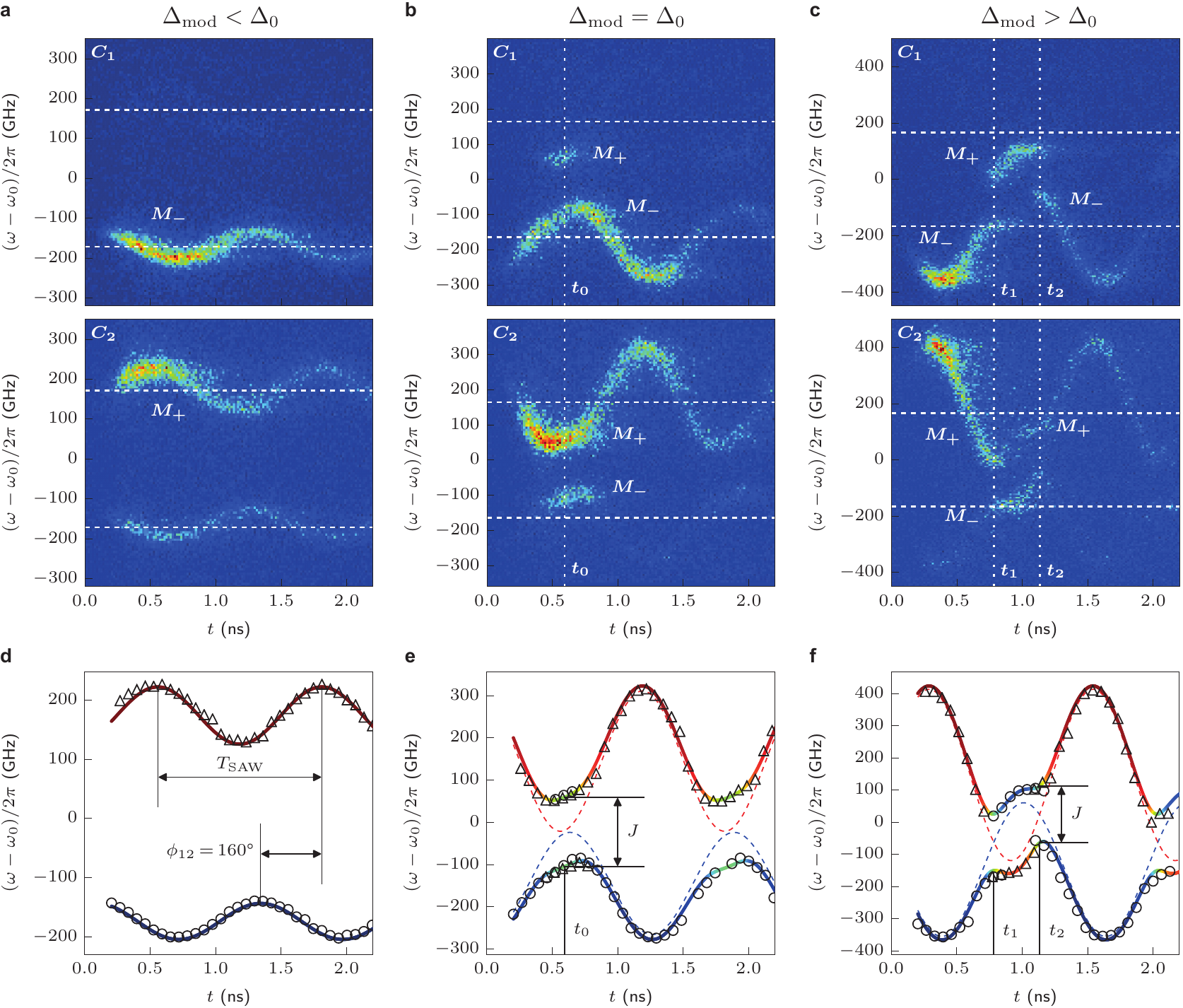}
\caption{
Experimental characterization for three different modulation amplitudes.
(a--c) Time resolved PL maps, measured on cavity $C_1$ (upper panels) and $C_2$ (lower panels).
(d--f) Mode frequencies extracted from the measurements on $C_1$ (circles) and $C_2$ (triangles), fitted with the coupled mode model.
For $\Delta_\mathrm{mod}<\Delta_0$ (a and d), the modes are effectively decoupled.
Their frequencies are modulated sinusoidally by the SAW with a phase difference of $\phi_{12}=160\textrm{\textdegree}$.
For $\Delta_{\mathrm{mod}}=\Delta_0$ (b and e), there is a time $t_0$ at which the cavities are in resonance.
At this point the modes are split by the coupling strength $J$ and they are delocalized over the photonic molecule, thus, both modes can be observed simultaneously on each cavity.
When $\Delta_{\mathrm{mod}}>\Delta_0$ (c and f), the modes exhibit an avoided crossing.
This happens twice in each SAW cycle, at times $t_1$ and $t_2$.
}
\label{fig2}
\end{figure*}

To experimentally verify the dynamic control of a PM by a $\omega_{\rm SAW}/2\pi=800\,{\rm MHz}$ SAW we performed stroboscopic spectroscopy which ensures optical excitation at a well-defined time during the acoustic cycle.
The emission of the PM was analysed in the time-domain and as a function of relative emission frequency $\omega-\omega_0$.
We measured time dependent emission spectra of the two cavities of PM1 for three characteristic modulation amplitudes.
For a weak modulation, $\Delta_{\rm mod}<\Delta_0$, shown in Fig. \ref{fig2} (a), $C_1$ and $C_2$ show the expected phase-shifted sinusoidal spectral modulations centred (dashed horizontal lines) at their unperturbed resonances which decay with a characteristic time constant of $\sim 1.3\,{\rm ns}$.
As the modulation is increased to $\Delta_{\rm mod}=\Delta_0$ (Fig. \ref{fig2} (b)) the two single cavity modes are brought into resonance at one distinct time $t_0 = 0.6\ \mathrm{ns}$ during the acoustic cycle.
At this time, coherent coupling leads to the formation of bonding and antibonding normal modes.
This directly manifests itself in the experimental data due to the emergence of new emission features stemming from the spatial delocalization of the bonding and antibonding modes.
For the initially lower frequency cavity $(C_1)$, a new signal appears at the frequency of the normal mode $M_+$, which was initially confined within the other cavity $(C_2)$.
The initially higher frequency cavity $(C_2)$ exhibits precisely the opposite behaviour, with the normal mode $M_-$ appearing at time $t_0$.
The normal modes are split by the coupling strength $J/2\pi=164\pm10\textrm{\,GHz}$, very close to the value expected from our FDTD simulations.
For further increased detuning $\Delta_{\rm mod}>\Delta_0$, the two modes are brought into resonance at two distinct times, $t_1$ and $t_2$, during the acoustic cycle.
After the first resonance at $t_1$, coupling is suppressed, $\Delta>J$, both modes are effectively decoupled and their single cavity characters are exchanged compared to the initial configuration.
The lower frequency mode $M-$, which is initially $C_1$-like, is switched to $C_2$-like character after $t_1$, and vice versa.
At the second resonance at $t_2$, the system is reverted to its original configuration at the beginning of the acoustic cycle.
This sequence of two time-offset coupling events gives rise to the experimentally observed anticorrelated time-evolutions of the emission of $C_1$ and $C_2$.
We extracted the time-dependent emission frequencies detected from the two cavities of PM1 and plotted them ($C_1$ circles, $C_2$ triangles) for the three modulation amplitudes in Fig. \ref{fig2} (d-f).
All experimental data are well reproduced by the simple coupled mode model given by equations (\ref{coupledmodes}) and (\ref{modulation}).
From the set of measurements with varying tuning amplitude we obtain mean values for the free parameters of the model, $\langle J\rangle_\mathrm{PM1}/2\pi=170\pm10\,\textrm{GHz}$, $\langle A_1/A_2\rangle_\mathrm{PM1}=0.75\pm0.05$ and $\langle\phi_{12}\rangle_\mathrm{PM1}=155\pm10\textrm{\textdegree}$.
Moreover, we note that the observed phase shift unambiguously confirms efficient coupling of the SAW into the sub-$\lambda_{\rm SAW}$ membrane.
Its small deviation from the ideal value of $180\textrm{\textdegree}$ arises from the finite difference in the phase velocity of the acoustic wave within the membrane and the region of the transducer.
The results from this model are plotted as lines and the character of the mode is color coded.
Clearly, our experimental data are in excellent agreement with the normal mode model for all three modulation amplitudes.
Both the time-dependent spectral modulation as well as the switching of the character of the modes are nicely reproduced.

\begin{figure}[htb]
\centering
\includegraphics[width=0.9\columnwidth]{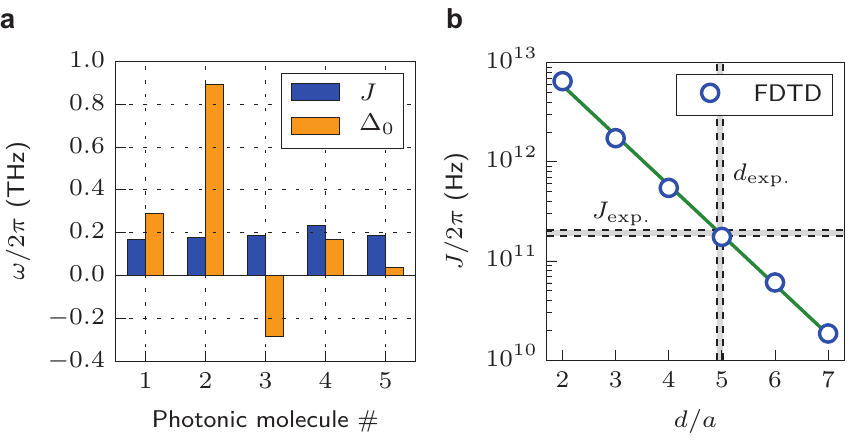}
\caption{Statistical variation of PM properties and comparison with FDTD simulation.
(a) Coupling strength $J$ and static detuning $\Delta_0$ measured on five nominally identical PMs.
(b) FDTD simulation of the coupling strength for different cavity separations $d$.
The experimentally determined coupling strength is accurately reproduced by the simulation.
}
\label{fig3}
\end{figure}

Such behaviour was experimentally confirmed for five different, nominally identical PMs for which we evaluated the mean of their respective key parameters.
The experimentally observed static detuning $\Delta_0$ and coupling strength $J$ are summarized in Fig. \ref{fig3} (a).
While the coupling strength, $J/2\pi = 192\pm 26\,{\rm GHz}$, (blue) does not vary from PM to PM, the values of the static detuning, $\Delta_0$, shows a pronounced scatter ranging between $-300\,{\rm GHz}$ and $+900\,{\rm GHz}$.
These observations are in fact expected.
$J$ exponentially depends on the symmetric inter-cavity offset, $d$, and is thus robust and insensitive to small deviations from the ideal geometry due to fabrication imperfections.
In strong contrast, the absolute resonance frequencies of the two cavities forming the PM are highly sensitive to these inherent and inevitable deviations from the nominal geometry.
The resulting fluctuations of the cavity resonances reflect themselves in the observed pronounced variation of $\Delta_0$.
Indeed, we observe pronounced coupling effects $(|J|>|\Delta_0|)$ only for one single as-fabricated PM, labeled PM5.
The corresponding experimental data is presented in the Supplementary information.
The exponential dependence of $J$ as a function of $d$ is nicely confirmed by a best fit (line) to values calculated by FDTD (symbols) presented in Fig. \ref{fig3} (b).
The experimentally observed distribution of $J_\mathrm{exp}$ and the inter-cavity separation $d_\mathrm{exp}$ derived from $J_\mathrm{exp}$ and the FDTD simulation are indicated by the shaded horizontal and vertical bars.
Clearly, the measured $J_\mathrm{exp}$ and its derived $d_\mathrm{exp}$ matching perfectly the calculated value of $J_\mathrm{sim}/2\pi=170\,{\rm GHz}$ and the nominally set $d=5a$.
These narrow distributions centred around the calculated and nominal values are expected since $d$ is large compared to typical imperfections in the nanofabrication.\\

In summary, we demonstrated dynamic optomechanical control of coherent interactions in a prototypical coupled nanophotonic system.
When combined with optical non-linearities our tunable PM paves the way to dynamically controlled high-fidelity entanglement generation \cite{Liew_2010} and distribution on a chip \cite{Cirac_1997,Vasco_2014}.
For larger switching rates as required for Landau-Zener-transition based gates \cite{Blattmann_2014}, the underlying optomechanical coupling could be enhanced further by direct SAW excitation of localized vibronic modes of PCM nanocavities \cite{Gavartin_2011} or shaped SAW waveforms \cite{Schuelein_2015}.
Finally we note, that our approach could be directly scaled up to large arrays of coupled cavities \cite{Notomi_2008,Liew_2013} or employed in superconducting two-level systems, which have recently been strongly coupled to single SAW quanta \cite{Gustafsson_2014}.

\section*{Methods}

\subsection*{Sample structure.}
We start by fabricating the photonic crystal membranes from a semiconductor heterostructure grown by molecular beam epitaxy.
 This heterostructure consists of a 170 nm GaAs layer with self-assembled InGaAs quantum dots (QDs) at its centre, on top of a 725 nm thick ${\rm Al_{0.8}Ga_{0.2}As}$ sacrificial layer.
The PM structure is defined by electron beam lithography and transferred into the heterostructure by ICP-RIE etching.
In a wet chemical etching step using hydrofluoric acid we removed the sacrificial layer to release a fully suspended membrane.
The PMs are deliberately designed to be off-resonant with the QD emission to achieve a sufficiently long decay time of the cavity emission.
Thus, we can observe spectral modulations over more than one acoustic cycle in the time domain, which would be impeded by Purcell-enhanced resonant QD exciton recombinations \cite{Fuhrmann_2011}.
IDTs were defined using electron beam lithography and metallized with 5 nm Ti and 50 nm Al in a lift-off process.
The finger period is $3.83\,\textrm{\textmu m}$, resulting in a resonance frequency of 800~MHz at 5~K.

\subsection*{Optical spectroscopy.}

For stroboscopic photoluminescence spectroscopy of the PM, off-resonant QDs are excited by an externally triggered diode laser emitting ~90 ps pulses at a wavelength of 850 nm which is focused to a $1.5\,\textrm{\textmu m}$ spot by a NIR $50\times$ microscope objective.
The emission is dispersed by a 0.75 m imaging grating monochromator.
A liquid ${\rm N_2}$-cooled Silicon charge coupled device and a fast ($< 50\,{\rm ps}$ rise time) Si-single photon avalanche detector (SPAD) are used for time-integrated multi-channel or time-resolved single channel detection, respectively.
The sample is cooled to $T = 5\,{\rm K}$ in a Helium-flow cryostat with custom built integrated rf connections.

\subsection*{Coupled mode model.}

We treat the PM as a model system of two coupled cavities.
The single cavity modes $M_{1,2}$ have the complex amplitudes $a_{1,2}$ and the frequencies $\omega_{1,2}$.
In the presence of finite coupling and the absence of dissipation, the time evolution is given by
\begin{equation*}
\begin{aligned}
\frac{da_1}{dt}=i\omega_1a_1+i\frac{J}{2}a_2\\
\frac{da_2}{dt}=i\omega_2a_2+i\frac{J}{2}a_1
\end{aligned}
\end{equation*}

with a real coupling constant $J$.
The resulting normal modes have the frequencies
$\omega_\pm=\omega_0\pm\frac{1}{2}\sqrt{\Delta^2+J^2}$ [Equation (\ref{coupledmodes})]
with centre frequency $\omega_0=\frac{1}{2}(\omega_1+\omega_2)$ and detuning $\Delta=\omega_2-\omega_1$.
We note that since $J>0$ the low frequency mode $\omega_-$ and the high frequency mode $\omega_+$ correspond to symmetric and antisymmetric superpositions of the uncoupled, single cavity modes.

\section*{Acknowledgements}
This work was supported by the Deutsche Forschungsgemeinschaft (DFG) via the Emmy-Noether-Programme (KR 3790/2-1), Sonderforschungsbereich 631 and the Cluster of Excellence \textit{Nanosystems Initiative Munich} (NIM). K. M. acknowledges support by the Alexander-von-Humboldt-Foundation.

\section*{Author contributions}
H.J.K. and S.K. designed study. S.K. built the experimental setup and performed the experiments. S.K., T.R. and S.L. designed, fabricated and characterised the devices, and performed 3D-FDTD simulations. K.M. fabricated and characterized the MBE material. S.K. and H.J.K. performed the data analysis and modelling. All authors discussed the results. S.K. and H.J.K. wrote the manuscript with contributions from all other authors. H.J.K., M.K., J.J.F. and A.W. inspired and supervised the project.

\end{document}